\documentclass[12pt]{article}
\pdfoutput=1
\usepackage[a4paper]{geometry}
\usepackage{jheppub, amsmath,amssymb,amsfonts,amsxtra, mathrsfs, makeidx,graphics,graphicx,amsthm,epsfig, ytableau,bm, longtable,float, color,tikz,mathtools,xfrac,footnote,rotating, lscape}
\usetikzlibrary{positioning}
\usetikzlibrary{chains}
\usetikzlibrary{arrows,fit, decorations.pathreplacing}
\definecolor{darkgreen}{rgb}{0.0, 0.2, 0.13}

\usepackage{bbm}

\numberwithin{equation}{section}

\newcommand{\nn}{\nonumber}

\newcommand{\be}{\begin{equation}} \newcommand{\ee}{\end{equation}}
\newcommand{\bea}{\begin{equation} \begin{aligned}} \newcommand{\eea}{\end{aligned} \end{equation}}

\def\hat{\widehat}

\def\bar{\overline}

\def\rt2{\sqrt{2}}

\def\Tr{\mathop{\rm Tr}}
\def\tr{\mathop{\rm tr}}

\def\CC{{\cal C}}

\def\CH{{\cal H}}

\def\CN{{\cal N}}

\def\CT{{\cal T}}

\def\1{{\ds 1}}

\def\repa{\raise4pt\hbox{$\square$}\mkern-14mu\raise-4pt\hbox{$\square$}}
\def\repab{\overline{\raise4pt\hbox{$\square$}\mkern-14mu\raise-4pt\hbox{$\square$}\mkern-1mu}}

\def\smileface{\ensuremath{\hbox{\large$\bigcirc$}\mkern-15mu\raise-1pt\hbox{\scriptsize$\smallsmile$}%
\mkern-10mu\raise4pt\hbox{..}\mkern4mu}}
\def\frownface{\ensuremath{\hbox{\large$\bigcirc$}\mkern-15mu\raise-1pt\hbox{\scriptsize$\smallfrown$}%
\mkern-10mu\raise4pt\hbox{..}\mkern4mu}}

\def\node#1#2{\overset{#1}{\underset{#2}{\circ}}}
\def\sqnode#1#2{\overset{#1}{\underset{#2}{{\color{gray} \blacksquare}}}}
\def\rnode#1#2{\overset{#1}{\underset{#2}{{\color{red} \bullet}}}}
\def\bnode#1#2{\overset{#1}{\underset{#2}{{\color{blue} \bullet}}}}
\def\gnode#1#2{\overset{#1}{\underset{#2}{{\color{gray} \bullet}}}}

\def\sqbnode#1#2{\overset{#1}{\underset{#2}{{\color{blue} \blacksquare}}}}
\def\sqrnode#1#2{\overset{#1}{\underset{#2}{{\color{red} \blacksquare}}}}

\def\bver#1#2{\overset{{\llap{$\scriptstyle#1$}\displaystyle{\color{blue} \bullet}{\rlap{$\scriptstyle#2$}}}}{\scriptstyle\vert}}
\def\ver#1#2{\overset{{\llap{$\scriptstyle#1$}\displaystyle\circ{\rlap{$\scriptstyle#2$}}}}{\scriptstyle\vert}}
\def\wver#1#2{\overset{{\llap{$\scriptstyle#1$}\displaystyle{\square}{\rlap{$\scriptstyle#2$}}}}{\scriptstyle\vert}}

\def\redsqver#1#2{\overset{{\llap{$\scriptstyle#1$}\displaystyle{\color{red} \blacksquare}{\rlap{$\scriptstyle#2$}}}}{\scriptstyle\vert}}

\tikzstyle{every picture}+=[remember picture]
\tikzstyle{na} = [baseline=-.5ex]

\newcommand{\ba}{\begin{array}}
\newcommand{\ea}{\end{array}}
\newcommand{\bi}{\begin{itemize}}
\newcommand{\ei}{\end{itemize}}

\def\bea#1\eea{\allowdisplaybreaks \begin{align}#1\end{align}}
 \newcommand{\ben}{\begin{enumerate}}
\newcommand{\een}{\end{enumerate}}
\newcommand{\bean}{\begin{eqnarray*}}
\newcommand{\eean}{\end{eqnarray*}}
\newcommand{\eref}[1]{(\ref{#1})}

\newcommand{\PE}{\mathop{\rm PE}}
\newcommand{\HS}{\mathop{\rm HS}}
\newcommand{\PL}{\mathop{\rm PL}}

\newcommand{\BC}{\mathbb{C}}

\newcommand{\BZ}{\mathbb{Z}}

\newcommand{\BH}{\mathbb{H}}

\newcommand{\comment}[1]{}

\newcommand{\diag}{\mathrm{diag}}

\newcommand{\purple}{\color{purple}}
\newcommand{\blue}{\color{blue}}

\newcommand{\red}{\color{red}}

\title{The Small $E_8$ Instanton and the Kraft Procesi Transition}
\author[a]{Amihay Hanany}
\author[b,c]{and Noppadol Mekareeya}

\affiliation[a]{Theoretical Physics Group, Imperial College London, \\
	Prince Consort Road, London, SW7 2AZ, UK}
\affiliation[b]{INFN, sezione di Milano-Bicocca, \\Piazza della Scienza 3, I-20126 Milano, Italy}
\affiliation[c]{Dipartimento di Fisica, Universit\`a di Milano-Bicocca, \\ Piazza della Scienza 3, I-20126 Milano, Italy}

\emailAdd{a.hanany@imperial.ac.uk}
\emailAdd{n.mekareeya@gmail.com}

\preprint{
	{\small
		\begin{flushright}
			IMPERIAL-TP-18-AH-01\\
		\end{flushright}
	}
}

\abstract{One of the simplest $(1,0)$ supersymmetric theories in six dimensions lives on the world volume of one M5 brane at a $D$ type singularity $\BC^2/D_k$. The low energy theory is given by an SQCD theory with $Sp(k-4)$ gauge group, a precise number of $2k$ flavors which is anomaly free, and a scale which is set by the inverse gauge coupling. The Higgs branch at finite coupling $\CH_f$ is a closure of a nilpotent orbit of $D_{2k}$ and develops many more flat directions as the inverse gauge coupling is set to zero (violating a standard lore that wrongly claims the Higgs branch remains classical). The quaternionic dimension grows by $29$ for any $k$ and the Higgs branch stops being a closure of a nilpotent orbit for $k>4$, with an exception of $k=4$ where it becomes $\bar{{\rm min}_{E_8}}$, the closure of the minimal nilpotent orbit of $E_8$, thus having a rare phenomenon of flavor symmetry enhancement in six dimensions. Geometrically, the natural inclusion of $\CH_f \subset \CH_{\infty}$ fits into the Brieskorn Slodowy theory of transverse slices, and the transverse slice is computed to be $\bar{{\rm min}_{E_8}}$ for any $k>3$. This is identified with the well known small $E_8$ instanton transition where 1 tensor multiplet is traded with 29 hypermultiplets, thus giving a physical interpretation to the geometric theory. By the analogy with the classical case, we call this the Kraft Procesi transition.
}
\begin{document}
\maketitle

\section{Anomaly coefficients and dynamical phenomena}
One of the simplest conditions for gravitational anomaly cancellation in 6d $\CN=(1,0)$ supersymmetric theories \cite{Green:1984bx, Sagnotti:1992qw} requires that for a theory with $n_V$ vector multiplets, $n_H$ hypermultiplets, and $n_T$ tensor multiplets, we need the relation \cite{RandjbarDaemi:1985wc,Dabholkar:1996zi}
\be
\label{6d29H1T}
n_H+29 n_T-n_V = 273 .
\ee
In dynamical processes, where the massless field content of the theory changes, this condition needs to be satisfied, even though the numbers $n_H, n_V, n_T$ may change. One should also note that when gravity is decoupled, the number 273 can change to another integer which is constant under the renormalisation group flow.
One of the simplest such dynamical phenomena is well known to be the supersymmetric Higgs mechanism, where the two quantities $n_T$ and $n_H-n_V$ remain fixed, while $n_H$ and $n_V$ take very interesting values depending on the gauge group and representation content of the low energy theory\footnote{There exist other examples of Higgs branch flows in 6d in which $n_T$ , $n_H$ , and $n_V$ all change. An example of this is a theory of $N$ M5-branes on $\BC^2/D_k$ discussed in section \ref{sec:M5Dk} of this paper. The number of tensor multiplets at a generic point of the tensor branch is $n_T= 2N-1$, whereas there are $N-1$ tensor multiplets left at the end of the Higgs branch flow \cite{Mekareeya:2017sqh}.  In this theory, $29n_T+n_H-n_V$ is equal to $30(N-1)+\dim(SO(2k))+1$, which is proportional to the anomaly coefficient $\delta$ given by \eref{anompolyDk}.}. This is a well studied phenomenon, is used in many physical systems in general, and in particular below, to evaluate dimensions of Higgs branches.  (The reader is referred to, for example, \cite{Cordova:2015fha, Mekareeya:2016yal, Mekareeya:2017sqh} for a recent development on the Higgs branch dimension of 6d $\CN=(1,0)$ theories.)

A much less studied phenomenon, which certainly deserves full attention, is the phenomenon of the small instanton transition, which was first pointed out in \cite{Ganor:1996mu} (see also \cite{Seiberg:1996vs, Intriligator:1997kq, Blum:1997mm, Hanany:1997gh} and \cite{Mekareeya:2017sqh} for subsequent work), where the quantities $n_V$ and $29n_T+n_H$ remain fixed while the numbers $n_H$ and $n_T$ change values. The simplest case is when one tensor multiplet is traded by 29 hypermultiplets and is known as the small $E_8$ instanton transition. It has the same dynamical significance as the supersymmetric Higgs mechanism, as massless fields change in numbers, while the system moves from one phase to another. We will show in this paper that also geometrically these two phenomena fit together into the theory of transverse slices.

This paper is devoted to the study of Higgs branches at finite and infinite coupling in a particular 6d $\CN=(1,0)$ gauge theory. The inverse coupling in such theories serves as a tension of a BPS string, and when it is tuned to zero, tensionless strings show up in the spectrum. Intuition from fundamental string physics suggests that infinitely many states become massless with the tensionless string, but a careful observation shows that most of these states are composites, and the number of states which are needed to characterize the theory are finite. In the language of Higgs branch, we say that the ring of BPS operators contains infinitely many additional operators, but they are finitely generated by a small set of operators which satisfy some non-trivial relations.

As the inverse gauge coupling is tuned to zero, the massless modes of the tensionless string can admit vacuum expectation values and new flat directions open up on the Higgs branch. The new moduli give mass to the tensor multiplet and consistency with Equation \eref{6d29H1T} requires that precisely 29 additional massless hypermultiplets should show up in the spectrum, providing the additional flat directions.

Geometrically, the Higgs branch grows in dimension by 29, and our goal is to determine the precise form of $\CH_\infty$, the Higgs branch at infinite coupling. Luckily, a 3d Coulomb branch allows for this evaluation and we find that the new massless states at the tensionless string limit transform in the spinor representation of the global symmetry, while all other massless states are composites of these new states, together with the states that already generate the Higgs branch at finite coupling -- mesons present in the IR theory and transform in the adjoint representation of the global symmetry.

Let us begin by describing the gauge theory and some of its features.

\section{One M5-brane on $D_k$ singularity}
\label{sectionM5}
The worldvolume theory of 1 M5 brane on a $D_k$ singularity is a 6d $\CN=(1,0)$ theory with one tensor multiplet and a gauge group $USp(2k-8)$ with $2k$ flavours \cite{Intriligator:1997kq,  Brunner:1997gk,Blum:1997mm, Intriligator:1997dh, Brunner:1997gf, Hanany:1997gh, Ferrara:1998vf, DelZotto:2014hpa}.  We denote this theory by the following quiver
\be \label{6doneM5}
\bnode{}{2k-8} - \sqrnode{}{4k}
\ee
where the blue node with an even integer label $m$ denotes the group $USp(m)$ and the red node with an integer label $n$ denotes the group $O(n)$.  The finite coupling Higgs branch $\CH_f$ of this theory, using the $F$ and $D$ term equations, is given by the set of all $4k\times 4k$ antisymmetric matrices $M$ with complex entries that square to 0 and has rank at most $2k-8$:
\be
\label{finiterel}
\CH_f = \left\{M_{4k\times 4k}| M+M^T=0, M^2=0, r(M)\le 2k-8\right\}~.
\ee 
This gives an algebraic description of the closure of the nilpotent orbit of $SO(4k)$ of height\footnote{The characteristic height is defined in (C.11) of \cite{Hanany:2017ooe}.} 2, corresponding to the partition $[2^{2k-8},1^{16}]$ of $4k$\footnote{In 3d, the quiver \eref{6doneM5} corresponds to $T_{[2k+7, 2k-7]}(SO(4k))$ in the notation of \cite{Gaiotto:2008ak}.}. It has dimension
\be \label{dimf6d}
\begin{split}
	\dim_\BH ~\CH_f[\eref{6doneM5}]  &= n_H[\eref{6doneM5}]  - n_V[\eref{6doneM5}]   \\
	&=  2k(2k-8)- \frac{1}{2}(2k-8)(2k-7) \\
	&= (k-4)(2k+7) = 2k^2-k-28~.
\end{split}
\ee
Due to the property of having height 2, $\CH_f$ has a particularly simple highest weight generating function (HWG)\footnote{The plethystic exponential (PE) of a multivariate function $f(x_1,x_2, \ldots, x_n)$ such that $f(0,0,\ldots, 0)=0$ is defined as $\PE[f(x_1,x_2, \ldots, x_n)] = \exp \left( \sum_{k=1}^\infty \frac{1}{k} f(x_1^k, x_2^k,\ldots, x_n^k) \right)$.} \cite{Ferlito:2016grh, Hanany:2016gbz},

\be
\label{HWGheight2}
\PE \left [ \sum_{i=1}^{k-4} \mu_{2i}t^{2i}\right].
\ee
In particular for the special case of $k=4$, $\CH_f$ is trivial and for the case of $k=5$, $\CH_f$ is the closure of the minimal nilpotent orbit of $SO(20)$. 
For $k=6$, the Hilbert series takes the form
\be
		\begin{split}
			\HS_{k=6} \, \CH_f[\eref{6doneM5}]  &=
			\sum_{p_2=0}^\infty\sum_{p_4=0}^\infty [0, p_2, 0, p_4, 0, 0, 0, 0, 0, 0, 0, 0] t^{2p_2 +4 p_4}~,
		\end{split} 
\ee
and so on.

Now let us consider the Higgs branch at infinite coupling, denoted by $\CH_\infty$.  The dimension of this space is 
\be \label{diminf6d}
\begin{split}
	\dim_\BH ~\CH_\infty[\eref{6doneM5}]  &= 29 n_T[\eref{6doneM5}]  +n_H[\eref{6doneM5}]  - n_V[\eref{6doneM5}]   \\
	&= (29 \times 1) + 2k(2k-8)- \frac{1}{2}(2k-8)(2k-7) \\
	&= 2k^2-k+1~,
\end{split}
\ee
having a 29 dimensional increase from $\CH_f$. This is a typical case of the small $E_8$ instanton transition \cite{Ganor:1996mu} in which 1 tensor multiplet is traded with 29 hypermultiplets, as discussed in the previous section. For the special case of $k=4$, as $\CH_f$ is trivial, we expect to find two effects. First that there is an enhancement of the flavor symmetry from $SO(16)$ to $E_8$, in order to fit with the known effect of \cite{Ganor:1996mu}. Second we expect that $\CH_\infty$ is the minimal hyperK\"ahler cone that has an $E_8$ isometry, namely the closure of the minimal nilpotent orbit of $E_8$. Indeed, this space is well known to have a dimension equal to 29, which is in perfect agreement with these expectations. Any other space with an $E_8$ isometry is known to have a dimension which is strictly above 29, hence the knowledge of the dimension and the isometry fixes the space uniquely. The purpose of this paper is to extend the picture to $k>4$, and to get a good description of $\CH_\infty$.

There is a natural inclusion of $\CH_f\subset\CH_\infty$, which calls for the Brieskorn-Slodowy theory of transverse slices (for a simple exposition of this program see \cite{Cabrera:2017njm} and references therein). For the case of $k=4$ the transverse slice of $\CH_f$ inside $\CH_\infty$ is $S=\CH_\infty = \bar{{\rm min}_{E_8}}$ itself, as $\CH_f$ is trivial, but for $k>4$ the computation is not trivial. Luckily, the answer turns out to be independent of $k$ and realizes the 29 dimensional increase in the Higgs branch. We say that locally the variety $\CH_\infty$ looks like a direct product $S\times \CH_f$.
Thus we find a physical realization to the transverse slice in the form of the small $E_8$ instanton transition of \cite{Ganor:1996mu}. The analogy with the case where the isometry group is classical \cite{Cabrera:2016vvv,Cabrera:2017njm} suggests to include this small instanton transition under the general class of the Kraft Procesi (KP) transitions \cite{Kraft1982}. In these cases, the transition follows from a simple Higgs mechanism, but the discussion following Equation \eref{6d29H1T}, and the geometric behavior suggests that these effects should be included under the same class.

It should be noted that while results on KP transitions are mostly known for Hasse diagrams of nilpotent orbits (see \cite{2015arXiv150205770F}, \cite{Chacaltana:2012zy} and \cite{Heckman:2016ssk} and references therein), the results of this paper show a nice extension to the case where the bigger moduli space is not a closure of a nilpotent orbit. The method which allows for the extension of the algebraic techniques of \cite{2015arXiv150205770F} uses a notion of ``difference" of quivers, as used successfully in \cite{Cabrera:2016vvv,Cabrera:2017njm}, and promises to generalize to many other cases that do not involve nilpotent orbits.

To facilitate the notion of ``difference" it is instructive to view the Higgs branch $\CH_f$ of quiver \eref{6doneM5} as a Coulomb branch of a 3d $\CN=4$ gauge theory.  There are two mirror theories of $USp(2k-8)$ gauge theory with $2k$ flavours.  
One can be obtained by using the brane construction involving an O3-plane (see \cite[Fig. 13]{Feng:2000eq}):
{\footnotesize
	\be \label{mirrUSp2km8w2kA}
	\rnode{}{2} - \bnode{}{2}-\rnode{}{4} - \bnode{}{4} - \cdots - \rnode{}{2k-10} - \bnode{}{2k-10}  - \rnode{}{2k-8}  -\bnode{\redsqver{}{1}}{2k-8} -  \underbrace{\rnode{}{2k-7}- \bnode{}{2k-8} - \cdots- \rnode{}{2k-7}}_{6\,\,\text{blue nodes} \, \, \& \,\, 7\,\,\text{red nodes}}- \bnode{\redsqver{}{1}}{2k-8}  - \rnode{}{2k-8}  - \bnode{}{2k-10} - \rnode{}{2k-10}- \cdots -\bnode{}{4} - \rnode{}{4} -\bnode{}{2} - \rnode{}{2}
	\ee}
The other can be obtained by using the brane construction involving an O5-plane (see \cite[sec. 4.1.1 \& Fig. 12]{Hanany:1999sj}. This quiver shows up in \cite{2013arXiv1309.0572H} in the study of Slodowy slices.):
\be \label{mirrUSp2km8w2kB}
\node{}{1}- \node{}{2}-\cdots -\node{}{2k-9}-\underbrace{\node{\wver{}{\,\,1}}{2k-8} - \node{}{2k-8} - \ldots - \node{\ver{}{k-4}}{2k-8}}_{7~\text{\comment{$(2k-8)$} nodes}}-\node{}{k-4}~, 
\ee

The Coulomb and Higgs branch dimensions of \eref{mirrUSp2km8w2kA} are
\be \label{dimCHmirrUsp2km8w2kA}
\begin{split}
	\dim_\BH \CC[\eref{mirrUSp2km8w2kA}]  &=  2 \times 2\left[ \sum_{j=1}^{k-4} j \right]+6(k-4) +7(k-4)= 2k^2-k-28\\
	\dim_\BH \CH[\eref{mirrUSp2km8w2kA}]  &= 2 \times \frac{1}{2} \left[ \sum_{j=1}^{k-4} (2j)^2 + \sum_{j=1}^{k-5} (2j)(2j+2)\right] + 14 \times \frac{1}{2}(2k-8)(2k-7)+2 \times \frac{1}{2}(2k-8) \\
	& \qquad  - 2 \left[ \sum_{j=1}^{k-5} \frac{1}{2}(2j)(2j+1) \right] - 2 \left[ \sum_{j=1}^{k-4} \frac{1}{2}(2j)(2j-1) \right]  \\
	& \qquad -2 \times \frac{1}{2}(2k-8)(2k-7) - 6 \times \frac{1}{2}(2k-8)(2k-7)-7 \times \frac{1}{2}(2k-7)(2k-8) \\
	& = k-4
\end{split}
\ee

While the Coulomb and Higgs branch dimensions of \eref{mirrUSp2km8w2kB} are the same,
\be \label{dimCHmirrUsp2km8w2k}
\begin{split}
	\dim_\BH \CC[\eref{mirrUSp2km8w2kB}]  &= \left[ \sum_{j=1}^{2k-9} j \right]+7(2k-8)+2(k-4) = 2k^2-k-28\\
	\dim_\BH \CH[\eref{mirrUSp2km8w2kB}]  &= \left[\sum_{j=1}^{2k-9} j(j+1) \right] +(2k-8)+6(2k-8)^2+2(k-4)(2k-8)\\ 
	& \quad -\left[ \sum_{j=1}^{2k-9} j^2 \right] - 7(2k-8)^2-2(k-4)^2 \\
	& = k-4
\end{split}
\ee
They are equal to the Higgs and Coulomb branch dimensions of $USp(2k-8)$ gauge theory with $2k$ flavours, respectively. Explicit computation shows that not only the dimensions coincide, but also the moduli spaces are the same, but we will not deal with this computation here.

Let us quote some important physical quantities which are used below -- the anomaly coefficients. The anomaly polynomial for the theory \eref{6doneM5} is explicitly given by \cite[(3.19)]{Ohmori:2014kda} (see also \cite{Intriligator:2014eaa, Cordova:2015fha}):\footnote{Note that
$\tr_{\text{adj}} F^2 = h^\vee_{SO(2k)} \tr_{\text{fund}} F^2$ and $\tr_{\text{adj}} F^4 = (2k-8) \tr_{\text{fund}} F^4 + 3 (\tr_{\text{fund}} F^2)^2$.}
\be \label{anompoly}
\begin{split}
I_8 &= \alpha c_2(R)^2 + \beta c_2(R) p_1(T) + \gamma p_1(T)^2 + \delta p_2(T) \\
& \qquad + \frac{1}{h^\vee_{SO(2k)}}\left( -\frac{x}{8} c_2(R) + \frac{h^\vee_{SO(2k)}}{96} p_1(T) \right)( \tr_{\text{adj}} F_L^2 + \tr_{\text{adj}} F_R^2) \\
& \qquad + \frac{1}{48} (\tr_{\text{adj}} F^4_L+ \tr_{\text{adj}} F_R^4) - \frac{1}{2(h^\vee_{SO(2k)})^2} \left(\frac{1}{4} \tr_{\text{adj}} F_L^2 - \frac{1}{4} \tr_{\text{adj}} F^2_R \right)^2~,
\end{split}
\ee
where 
\bea
\alpha &= \frac{1}{24} |\Gamma_{D_k}|^2 - \frac{1}{12} \left[ |\Gamma_{D_k}| (k+1) -1 \right] + \frac{1}{24} [\dim(SO(2k)) -1]~,\\
\beta &= \frac{1}{48} \left[2-  |\Gamma_{D_k}|(k+1) \right] +  \frac{1}{48} [\dim(SO(2k)) -1]~, \\
\gamma &= \frac{7}{5760}  \left[ \dim(SO(2k)) +1 \right] \\
\delta &= -\frac{1}{1440} \left[  \dim(SO(2k)) +1 \right] \\
x &= |\Gamma_{SO(2k)}| - h^\vee_{SO(2k)}~.
\eea
and 
\be |\Gamma_{D_k}|=4k-8 \ee is the order of the dihedral group $\hat{D}_{k}$, which fits in the McKay correspondence with the group $SO(2k)$; the dimension of $SO(2k)$ is 
\be \dim(SO(2k)) = k(2k-1)~; \ee    
and the dual Coxeter number of $SO(2k)$ is
\be
h^\vee_{SO(2k)} = 2k-2~.
\ee

It is worth noting that, for a general 6d $\CN=(1,0)$ theory, the anomaly coefficient $\delta$ is related to the numbers of tensor multiplets $n_T$, vector multiplets $n_V$ and hypermultiplets $n_H$ by \cite{Cordova:2015fha}
\be
\delta = -\frac{1}{1440} (29n_T+n_H-n_V)~.
\ee

\subsection{$T^2$ and $T^3$ compactifications of the $6d$ theory} 
As proposed by \cite{Ohmori:2015pua}, the $T^2$ compactification of the SCFT at infinite coupling of \eref{6doneM5} is a 4d theory of class $\mathsf{S}$ of $SO(2k)$-type associated with a sphere with two maximal punctures $[1^{2k}]$ and one minimal puncture $[2k-3,3]$.\footnote{The convention for the labels of the punctures is in accordance with \cite{Chacaltana:2012zy}.}  We denote this theory by 
\be \label{classSoneM5}
\mathsf{S} \langle S^2 \rangle_{SO(2k)} \{ [1^{2k}],[1^{2k}], [2k-3,3] \} 
\ee
The flavour symmetries associated with the punctures $[1^{2k}]$ and $[2k-3,3]$ are $SO(2k)$ and trivial respectively.  Hence the flavour symmetry that is manifest in \eref{classSoneM5} is $SO(2k) \times SO(2k)$.  However, as pointed out in \cite[the 2nd row on p. 26]{Chacaltana:2011ze}, for $k=4$, the theory is identified with the rank-1 $E_8$ SCFT, whose flavour symmetry is $E_8$.  For $k \geq 5$, it can be checked, for example using the Hilbert series or the Hall-Littlewood index, that the theory \eref{classSoneM5} has an $SO(4k)$ flavour symmetry. These points are in agreement with our expectations above.

The central charges of theory \eref{classSoneM5} is\footnote{Let us follow the method described in \cite{Chacaltana:2011ze}. The effective number of vector multiplets is $n_V = -\frac{1}{3} k (16 k^2 - 24 k + 11) +2 \delta n^\text{(max)}_V + \delta n^\text{(min)}_V$ and the effective number of hypermultiplets is $n_H = -\frac{8}{3} k (k - 1) (2 k - 1)+2 \delta n^\text{(max)}_H + \delta n^\text{(min)}_H$, where $\delta n^\text{(max)}_H = \frac{4}{3} (k-1) k (2 k-1)$, $\delta n^\text{(max)}_V =\frac{1}{3} (k-1) k (8 k-7)$, $\delta n^\text{(min)}_H = 4 k^2-4 k-8$ and $\delta n^\text{(min)}_V =4 k^2-4 k-9$.  The central charges are therefore $a = \frac{5}{24} n_V + \frac{1}{24} n_H =\frac{7 k^2}{12}-\frac{19 k}{24}-\frac{53}{24}$ and  $c = \frac{1}{6} n_V + \frac{1}{12} n_H= \frac{2 k^2}{3}-\frac{5 k}{6}-\frac{13}{6}$.}
\be \label{centralchargesSO}
a= \frac{7 k^2}{12}-\frac{19 k}{24}-\frac{53}{24}~, \qquad c= \frac{2 k^2}{3}-\frac{5 k}{6}-\frac{13}{6}~.
\ee
These can indeed be obtained from the anomaly coefficients in \eref{anompoly} as follows \cite[(6.4)]{Ohmori:2015pua}:
\be
a= 24 \gamma -12 \beta - 18 \delta~, \qquad c= 64 \gamma -12 \beta - 8 \delta~.
\ee 

If we compactify \eref{classSoneM5} further on $S^1$ and use 3d mirror symmetry, we obtain the following star-shaped quiver
\be \label{SOSpmirr}
\rnode{}{2} - \bnode{}{2}-\rnode{}{4} - \bnode{}{4} - \cdots - \rnode{}{2k-2} - \bnode{}{2k-2} - \rnode{\bver{}{2}}{2k} - \bnode{}{2k-2}- \rnode{}{2k-2} - \cdots  - \bnode{}{4}-\rnode{}{4}  - \bnode{}{2}- \rnode{}{2}
\ee
The Coulomb branch dimension of \eref{SOSpmirr} is
\be \label{SOSpmirrCoul}
\begin{split}
	\dim_\BH \,\CC[\eref{SOSpmirr}] 
	= 4\sum_{j=1}^{k-1} j + k+1 = 2k^2-k+1 = \eref{diminf6d}~.
\end{split}
\ee
On the other hand, the Higgs branch dimension of \eref{SOSpmirr} is
\be  \label{SOSpmirrHiggs}
\begin{split}
	\dim_\BH \,\CH[\eref{SOSpmirr}] 
	&= 2 \sum_{j=1}^{k-1} \frac{1}{2}(2j)(2j) +2 \sum_{j=1}^{k-1} \frac{1}{2}(2j)(2j+2) +2k \\
	&\qquad  - \sum_{k=1}^k \frac{1}{2}(2j)(2j-1) - \sum_{k=1}^{k-1} \frac{1}{2}(2j)(2j-1)  \\
	&\qquad - \sum_{k=1}^{k-1} \frac{1}{2}(2j)(2j+1) -3 \\
	&=  k-3 \\
	&=  n_T[\eref{6doneM5}] + (\text{the rank of the gauge group in \eref{6doneM5}})~,
\end{split}
\ee
where $n_T[\eref{6doneM5}]=1$ and {the rank of the gauge group in \eref{6doneM5} is $k-4$.  For the special case of $k=3$, we have $\dim_\BH \,\CH[\eref{SOSpmirr}] = 16$ and $\dim_\BH \,\CH[\eref{SOSpmirr}] = 0$. This indicates that for $k=3$, the corresponding 6d theory is the theory of 16 free hypermultiplets. Indeed, this is as expected for the theory of a single M5 brane on an $A_3$ singularity. This point is used below.

\subsubsection*{{\it An alternative description -- A minimally unbalanced quiver}}
By analogy with the finite coupling case in which there are two quiver descriptions, one with alternating $USp$/$SO$ groups, and one with unitary groups, let us propose another description of the 3d mirror theory of $T^3$ compactification of the SCFT at infinite coupling of \eref{6doneM5}:
\be \label{starSU}
\node{}{1}- \node{}{2}-\cdots -\node{}{2k-3}-\node{\ver{}{k-1}}{2k-2}-\node{}{k}-\gnode{}{2}~, 
\ee 
where each node labelled by $m$ denotes the $U(m)$ gauge group and the unbalanced node is denoted in grey, with an imbalance of $k-4$.  For $k=4$, \eref{starSU} is the affine $E_8$ Dynkin diagram, whose Coulomb branch is the reduced moduli space of one $E_8$ instanton or, equivalently $\bar{{\rm min}_{E_8}}$, the closure of the minimal nilpotent orbit of $E_8$.  This is in agreement with the Coulomb branch of \eref{SOSpmirr}, and with the expectations in Section \eref{sectionM5}.  Below we provide arguments on how we came up with the quiver \eref{starSU}. It should be noted that this family of quivers appears in five dimensional gauge theories as the Higgs branch of the UV fixed point \cite{Ferlito:2017xdq} of $SU(k-2)_{\pm \frac{1}{2}}$ gauge theory with $2k-1$ flavours \cite{Hayashi:2015fsa} (or $USp(2k-6)$ gauge theory with $2k-1$ flavors \cite{Gaiotto:2015una}). Comparing with the six dimensional theories studied in this paper, namely $USp(2k-8)$ with $2k$ flavours, we find a shift by 1 of the number of flavours. This shift can be explained by recalling the points raised in \cite{Tachikawa:2015mha, Hayashi:2015fsa, Yonekura:2015ksa, Zafrir:2015rga}, which argue that when the number of flavours is increased by one, namely for a 5d $\CN=1$ $SU(k-2)_0$ gauge theory with $2k$ flavours, the fixed point is in six dimensions while the flavour symmetry $SO(4k)$ remains the same as that of 5d $\CN=1$ $SU(k-2)_{\pm \frac{1}{2}}$ gauge theory with $2k-1$ flavours. To add to their point, we also expect that the Higgs branch of the 5d UV theory and the 6d UV theory are the same. The papers also propose that the symplectic gauge group in 6d theory is broken to the unitary gauge group in 5d theory by a Wilson line from the circle compactification. 
 

Let us proceed and point out that the Coulomb branch dimension is
\be \label{CoulstarSU}
\dim_\BH \, \CC[ \eref{starSU}]  = \left[ \sum_{j=1}^{2k-2} j \right] + (k-1)+k+2-1 = 2k^2-k+1 = \eref{SOSpmirrCoul}.
\ee
and the Higgs branch dimension is
\be
\begin{split}
	\dim_\BH \, \CH[ \eref{starSU}]  &= \left[\sum_{j=1}^{2k-3} j(j+1) \right] +(2k-2)(k-1) + k(2k-2) +2k  \\
	& \qquad - \left[ \sum_{j=1}^{2k-2} j^2 \right] - k^2 -(k-1)^2-4+1 \\
	&= k-3 = \eref{SOSpmirrHiggs}~.
\end{split}
\ee
Similarly to what has been discussed for $\eref{SOSpmirr}$, in the case of $k=3$, we have $\dim_\BH \, \CC[ \eref{starSU}] = 16$ and $\dim_\BH \, \CH[ \eref{starSU}] = 0$. This indicates that for $k=3$, the corresponding 6d theory is the theory of 16 free hypermultiplets, as expected for the worldvolume theory of a single M5 brane on the $A_3$ singularity.

Moreover, \eref{starSU} is the mirror theory of the $S^1$ compactification of the class $\mathsf{S}$ theory 
\be \label{classSoneM5SU}
\mathsf{S} \langle S^2 \rangle_{SU(2k-2)} \{ [1^{2k-2}],[(k-1)^{2}], [(k-2)^2,2] \}~, 
\ee
which has central charges\footnote{Let us follow the method described in \cite{Chacaltana:2010ks}.  The pole structures of the punctures $[1^{2k-2}]$, $[(k-1)^{2}]$ and $[(k-2)^2,2]$ are $p^{(1)} =\{1,2,3,\ldots, 2k-3\}$, $p^{(2)} = \{1,1,2,2,3,3, \ldots,k-2,k-2,k-1\} $, $p^{(3)} = \{1,2,2,3,4,4,5,5, \ldots, k-1,k-1, k\}$, respectively. The graded Coulomb branch dimensions $(d_2, d_3, \ldots, d_{2k-2})$ of this theory are $d_m = -(2m-1) +\sum_{i=1}^3 p^{(i)}_{m-1}$; hence, $d_2=d_3=d_4=d_5=0$, $d_{2\ell} =1$ and $d_{2\ell+1}=0$ for $\ell \geq 3$.  The effective number of vector multiplet is $n_V = \sum_{j=2}^{2k-2} (2j-1)d_j =  2 k^2-3k-9$, and the effective number of hypermultiplets is $n_H = [\text{the dimension of the Higgs branch of \eref{classSoneM5}}] + n_V = \eref{CoulstarSU} + n_V = 4 k^2-4 k-8$.  The central charges are therefore $a = \frac{5}{24} n_V + \frac{1}{24} n_H =\frac{7 k^2}{12}-\frac{19 k}{24}-\frac{53}{24}$ and  $c = \frac{1}{6} n_V + \frac{1}{12} n_H= \frac{2 k^2}{3}-\frac{5 k}{6}-\frac{13}{6}$.} 
\be
a= \frac{7 k^2}{12}-\frac{19 k}{24}-\frac{53}{24}~, \qquad c= \frac{2 k^2}{3}-\frac{5 k}{6}-\frac{13}{6}~.
\ee
These are in agreement with \eref{centralchargesSO}.  Hence we propose the following dualities:
\be
\begin{split}
	\eref{classSoneM5} \quad &\longleftrightarrow \quad \eref{classSoneM5SU} \\
	\eref{SOSpmirr} \quad &\longleftrightarrow \quad \eref{starSU}
\end{split}
\ee

Let us now provide further arguments for the quiver family \eref{starSU}, which follow the reasoning presented in \cite{Ferlito:2017xdq}. This family of quivers also appears in \cite{Hanany:2017pdx} in the context of symmetry enhancements and exact HWG's. The global symmetry at finite coupling is $SO(4k)$, and is assumed to remain the global symmetry at infinite coupling. This is in contrast to phenomena in 5d and in 3d where symmetry enhancements are rather central and frequent phenomena for theories with this amount of supersymmetry. In 3d the symmetry enhancement is caused by monopole operators of spin 1 under $SU(2)_R$, and in 5d it is caused by instanton operators, again with spin 1 under $SU(2)_R$. 6d global symmetry enhancements happen in rather rare cases, and it is therefore important to understand how they come about.

An inspection of the $k=4$ case shows one of these relatively rare cases where there is an enhancement of the global symmetry from $SO(16)$ to $E_8$. The rank is preserved and indeed $SO(16)$ is one of the largest Levy subgroups of $E_8$. The enhancement indicates that as we tune the inverse coupling to zero, there are extra massless states which transform in the spinor representation of $SO(16)$ and have spin 1 under $SU(2)_R$ (This uses the theorem by Kostant and Brylinski \cite{1992math......4227B} that operators with spin 1 under $SU(2)_R$ transform in the adjoint representation of the global symmetry). It is reasonable to assume that such states exist for higher $k$, transforming again under the spinor representation of $SO(4k)$, but with higher representation of $SU(2)_R$, hence not contributing to the symmetry enhancement. The simplest assumption is a linear dependence on $k$ giving spin $\frac{k-2}{2}$ under $SU(2)_R$. This behavior ensures that at $k=4$ the $SU(2)_R$ spin is 1 and at $k=3$ it is $\frac{1}{2}$, as expected from a $D_3=A_3$ singularity, which is known to give a free theory, as indicated above. We are led to a challenge of looking for such extra states in the gauge theory. Indeed, one can find these in the D2 D6 $O6$ brane system.

\be \label{brane}
\begin{tikzpicture}[baseline]
\filldraw (0,1.3) circle (0.2cm) node[xshift =0cm, yshift=0.4cm] {\tiny NS5};
\filldraw (2,1.3) circle (0.2cm) node[xshift =0cm, yshift=0.4cm] {\tiny NS5};
\draw [very thick] (0.2,1.24)--(1.8,1.24) node[black,midway, yshift=-0.3cm] {\scriptsize $(k-4)$ D6};
\draw [very thick] (2,1.24)--(4,1.24) node[black,midway, yshift=-0.3cm] {\scriptsize $k$ D6};
\draw [very thick] (-2,1.24)--(0,1.24) node[black,midway, yshift=-0.3cm] {\scriptsize $k$ D6};
\draw [darkgreen, dashed, very thick] (-2,1.28)--(4,1.28) node[black,xshift=0.5cm, yshift=0cm] {\scriptsize {\color{darkgreen} O6}} node[black,xshift=-5cm, yshift=0.2cm] {\scriptsize \color{darkgreen} $-$} node[black,xshift=-3cm, yshift=0.2cm] {\scriptsize\color{darkgreen} $+$} node[black,xshift=-1cm, yshift=0.2cm] {\scriptsize \color{darkgreen} $-$}
node[black,xshift=2cm, yshift=0cm] {\scriptsize $\overset{x^6}{\longrightarrow}$};
\draw [red, very thick] (0.2,1.28)--(1.8,1.28) node[black,xshift=-1.3cm, yshift=0.2cm] {\tiny \red D2};
\end{tikzpicture} 
\ee

Consider the Type IIA brane system, depicted in \eref{brane}, of $k-4$ D6 branes on an $O6^+$ planes stretched between two NS5 branes, and a set of $k$ semi infinite D6 branes on $O6^-$ in each side of the NS5 branes. The branes occupy the following directions:
\bea
\begin{array}{c|cccccccccc}
	\hline
	& 0 & 1 & 2 & 3 & 4 & 5 & 6 & 7 & 8 & 9 \\
	\hline                     
	\mathrm{D6} & \mathrm{X} & \mathrm{X} & \mathrm{X} & \mathrm{X} & \mathrm{X}  & \mathrm{X} &\mathrm{X} &~ &~ &~  \\
	\mathrm{NS5}&\mathrm{X}& \mathrm{X} & \mathrm{X} & \mathrm{X}  &\mathrm{X} & \mathrm{X} &~ &~ &~& ~  \\
	\mathrm{D2} &\mathrm{X}& \mathrm{X} &~ &~ &~ &~ &\mathrm{X} &~ &~& ~\\
	\hline
\end{array}
\eea
The gauge coupling is proportional to the distance between the two NS branes and is measured by a vacuum expectation value of a real scalar field in a tensor multiplet. It is also a tension of a BPS string which is represented by the D2 brane, denoted in red in \eref{brane}, which is stretched between the two NS5 branes and is parallel to the D6 branes. A D6 brane observer sees the D2 brane as a gauge instanton, which carries fermionic zero modes from the flavor branes. Quantization of these zero modes leads to massless states in the spinor representation of the flavor symmetry in the limit in which the inverse gauge coupling is zero (see, for example, \cite{Polchinski:1995df}). Incidentally, as discussed above, there are infinitely many such massless states, as we tune a tension of a string to zero, however, it turns out that the states in the spinor representation are the essential states. We return to this point below.

We thus have a flavor symmetry of $SO(4k)$ and an additional state in the spinor representation of $SO(4k)$. Remarkably these two conditions, together with the assumption of existence of a 3d quiver, are sufficient to come up with the quiver of \eref{starSU}. We first need the notion of an imbalance of a quiver node as the number of its flavors minus twice the number of its colors. We also recall from \cite{Gaiotto:2008ak, Ferlito:2017xdq} that the set of balanced nodes forms the Dynkin diagram of the global symmetry. Furthermore the number of $U(1)$ factors in the global symmetry is equal to the number of unbalanced nodes minus 1. As for the case under discussion there are no $U(1)$ factors in the global symmetry, we find only one unbalanced node in the quiver. We only need to figure out where to place it. This is easily done as the extra state in the spinor representation indicates that the unbalanced node should be attached to the spinor node.

To summarize, the quiver is made out of a set of $2k$ balanced nodes which form the Dynkin diagram of $SO(4k)$, and one unbalanced node that is attached to one of the spinor nodes. The first $2k-2$ nodes form an increasing set from 1 to $2k-2$, while the two spinor nodes get values $k-1$ and $k$ in order to keep the balancing condition of the $2k-2$ node and first spinor node. Finally, the second  spinor node is attached to a node 2 for any $k$, again in order to keep it balanced, thus deriving quiver \eref{starSU}. The last node has an imbalance of $k-4$ which is indeed confirming the expectation that the representation under $SU(2)_R$ is $\frac{k-2}{2}$. The special case of $k=4$ reproduces the affine Dynkin diagram of $E_8$, as expected, and the case $k=3$ reproduces a free theory with 16 hypermultiplets \cite{Ferlito:2017xdq}, again consistent with expectations from the $A_3$ singularity. This is a special case of 1 M5 brane on an $A_n$ singularity with $(n+1)^2$ free hypermultiplets. 

%

\subsection{Comparing the Higgs branches of \eref{6doneM5} at finite coupling and at infinite coupling}
Let us denote the Higgs branches of \eref{6doneM5} at finite coupling and at infinite coupling by $\CH_{f}$ and $\CH_\infty$ respectively.  From quiver \eref{6doneM5}, and Equation \eref{finiterel} we recall
\be \label{Higgsfinite}
\CH_f[\eref{6doneM5}] = \text{the closure of the nilpotent orbit $[2^{2k-8},1^{16}]$ of $SO(4k)$}.
\ee
Using mirror symmetry and the assumption that the compactification of a 4d theory to 3d does not change the Higgs branch, we obtain
\be
\CH_\infty[\eref{6doneM5}] = \CC[\eref{starSU}]
\ee

\subsubsection{The infinite coupling Higgs branch of \eref{6doneM5} from the Coulomb branch of \eref{starSU}}
Let us now concentrate on the last node on the right hand side of \eref{starSU}. it has an imbalance of $N_f-2N_c=k-4$. The lowest $SU(2)_R$ spin for a 3d monopole operator with non-zero fluxes associated to this last gauge node is $(k-2)/2$. Such a monopole transforms in the spinor representation of $SO(4k)$.
Thus the chiral ring at infinite coupling is generated by an $SO(4k)$ adjoint rep at $SU(2)_R$ spin-1 and an $SO(4k)$ spinor rep at $SU(2)_R$ spin-$(k-2)/2$. There are now a collection of techniques to evaluate the chiral ring, and the reader is referred to \cite[Sec.~4.2.2]{Hanany:2017pdx}. The resulting highest weight generating function is \cite{Hanany:2017pdx, Ferlito:2017xdq}
\be \label{HWGSO4np8}
\text{HWG of $\CC$[\eref{starSU}]}  = \PE \left[ \sum_{i=1}^{k-1} \mu_{2 i} t^{2 i}+ t^4 + \mu_{2 k} (t^{k-2} + t^{k}) \right]~.
\ee
A simple observation of this HWG reveals that the lattice of weights for this moduli space consists of the adjoint and one of the spinor representations, but not the other 2 sub lattices of $SO(4k)$. This situation resembles the case of the perturbative spectrum of the Heterotic $SO(32)$ string where the gauge group is sometime said to be $Spin(32)/\BZ_2$.

Equation \eref{HWGSO4np8} encodes the representation content of all half BPS operators on the Higgs branch of the 6d SCFT. Comparing this equation with \eref{HWGheight2} we find the extra representations which are formally represented by the polynomial
\be
\mu_{2 k-6} t^{2 k-6}+ \mu_{2 k-4} t^{2 k-4}+ \mu_{2 k-2} t^{2 k-2}+ t^4 + \mu_{2 k} (t^{k-2} + t^{k})
\ee
The most important representation is the spinor representation of $SO(4k)$ which is encoded by the monomial $\mu_{2 k} t^{k-2}$, indeed with spin $\frac{k-2}{2}$ under $SU(2)_R$, as expected above. This is the only additional generator of the ring of half BPS operators, together with the adjoint representation of $SO(4k)$ at spin 1 under $SU(2)_R$. All other infinitely many representations which arise at infinite coupling do not encode generators of the ring, and result from tensor/symmetric products of the generators. As above, we see that there are infinitely many states which become massless as the tension of the string is tuned to zero, however, the supersymmetry and the ring structure allow these states to be composed of the basic ones -- in the adjoint and spinor representations of $SO(4k)$, or perhaps $Spin(4k)/\BZ_2$.

The Hilbert series computed from \eref{HWGSO4np8}, and their plethystic logarithms, are explicitly written in \cite[(2.5)--(2.8)]{Ferlito:2017xdq}, with $k=n+2$ for $n=2,3,4$.  Since the plethystic logarithm contains the information about the generators of the chiral ring and their relations, it is worth mentioning here the features for general $k$.  The characters of the following representations appear with a positive sign: $[0,1,0,\ldots, 0]$ at order $t^2$ and $[0, \ldots, 0,1]$ at order $t^{k-2}$.  These imply the existence of the generators in such representations at the corresponding order (twice spin under $SU(2)_R$).  On the other hand, the relations are in the following representations: $[2,0,\ldots,0]$ at order $t^4$, $[1,0,\ldots,0,1,0]$ at order $t^k$, and at order $t^{2k-4}$ the relations can be divided into two cases: for even $k$, we have $\wedge^{4j} [1,0,\ldots, 0]$ for $j=0, 1, \ldots, (k-2)/2$, and for odd $k$ we have $\wedge^{4j+2} [1,0,\ldots,0]$ for $j=0, 1, \ldots, (k-3)/2$. We provide the explicit expressions for the plethystic logarithms\footnote{The plethystic logarithm of a multivariate function $f(x_1,...,x_n)$ such that $f(0,...,0)=1$ is
\bea
\PL[f(x_1,...,x_n)]=\sum^\infty_{k=1} \frac{1}{k} \mu(k) \log f(x^k_1,...,x^k_n)~, \nn
\eea
where $\mu(k)$ is the Moebius function.  The plethystic logarithm of the Hilbert series encodes generators and relations of the chiral ring.} for $k=7$ and $k=8$ below.
\bea
\PL_{k=7}  \, \CH_\infty[\eref{6doneM5}]  &= [0,1,0^{12}] t^2 - [2,0^{13}] t^4 + [0^{13},1] t^5 + [2,0^{13}] t^6 - [1,0^{11},1,0] t^7 \nn \\
& \qquad - [0,1,0^{12}]t^8 +([2,0^{12},1]+[1,0^{11},1,0])t^9  \\
& \qquad + (-[0,1,0^{12}]-[0^5,1,0^{8}]-[0^9,1,0^{4}]+[0,1,0^{12}]) t^{10}+ \ldots \nn \\
\PL_{k=8}  \, \CH_\infty[\eref{6doneM5}]  &= [0,1,0^{14}] t^2 - [2,0^{15}] t^4 + ([0^{15},1] +[2,0^{15}]) t^6 - [1,0^{13},1,0] t^8 \nn \\
& \qquad +([2,0^{14},1]+[1,0^{13},1,0]+[0,1,0^{14}])t^{10} \nn \\
& \qquad +(-2[0^{16}]-[0^3,1,0^{12}]-[0^7,1,0^{8}]-[0^{11},1,0^{4}] \nn\\
& \qquad \,\,\, -[0^{15},1] -[2,0^{14},1] -[0,1,0^{13},1]  - [1,0^{13},1,0]  \\
& \qquad \,\,\, - [3,0^{13},1,0] -[2,0^{15}])t^{12} + \ldots~. \nn
\eea

\subsection{Kraft Procesi and the small instanton transition} 
As can be seen from \eref{diminf6d} the Higgs branch dimension ``jumps'' up by $29$ when one goes from finite coupling to infinite coupling.  This phenomenon is known as the small instanton transition \cite{Ganor:1996mu, Seiberg:1996vs, Intriligator:1997kq, Hanany:1997gh}.  One can elegantly realise this phenomenon from the perspective of three dimensional gauge theories by taking the ``difference" of quiver \eref{starSU} and quiver \eref{mirrUSp2km8w2kB}, as introduced in \cite{Cabrera:2016vvv} and further developed in \cite{Cabrera:2017njm}.  

In particular, one can obtain \eref{starSU} from \eref{mirrUSp2km8w2kB} by the following steps, which realizes the transverse slice of one space in another using the quiver language:
\ben
\item  Remove the square node with label 1 from \eref{mirrUSp2km8w2kB}.
\item  Add the ranks of the nodes in the following $E_8$ quiver to the $2$nd, $3$rd, $\ldots$, $7$th $(2k-8)$ nodes as well as the two $(k-4)$ nodes in \eref{mirrUSp2km8w2kB}.
\be
\label{AffineE8quiver}
\node{}{1} - \node{}{2}- \node{}{3}- \node{}{4}- \node{}{5} - \node{\ver{}{3}}{6}-\node{}{4}-\node{}{2}
\ee
\item As a result, we arrive at quiver \eref{starSU}.
\een
This ``superposition'' of the two quivers can be illustrated as follows\footnote{A.~H.~ would like to thank Hiraku Nakajima for inspiring discussions in Banff when this point was realized.}:
\be \label{superposition}
\begin{array}{lccccccccc}
	&\node{}{1}- \node{}{2}-\cdots -\node{}{2k-9}-\node{}{2k-8} - &\node{}{2k-8} - &\node{}{2k-8}- &\node{}{2k-8}- &\node{}{2k-8}- &\node{}{2k-8} - &\node{\ver{}{k-4}}{2k-8}-&\node{}{k-4} & \\
	&+ \quad & \, \node{}{1} - &\node{}{2}- &\node{}{3}- &\node{}{4}- &\node{}{5} - &\node{\ver{}{3}}{6}- &\node{}{4}- &\node{}{2} \\
	\longrightarrow \quad &\node{}{1}- \node{}{2}-\cdots -\node{}{2k-9}-\node{}{2k-8} - &\node{}{2k-7}- &\node{}{2k-6}- &\node{}{2k-5}- &\node{}{2k-4}- &\node{}{2k-3} - &\node{\ver{}{k-1}}{2k-2}- &\node{}{k}- &\node{}{2}~, 
\end{array}
\ee

and represents the quiver form of the transverse slice. We say that the Coulomb branch of \eref{mirrUSp2km8w2kB} is embedded inside the Coulomb branch of \eref{starSU} with a transverse slice given by the Coulomb branch of \eref{AffineE8quiver}.

Moreover, one can also obtain \eref{SOSpmirr} from \eref{mirrUSp2km8w2kA} in a similar way, along the lines detailed in \cite{Cabrera:2017njm}.  We move the two square nodes (labelled by 1) towards each other in \eref{mirrUSp2km8w2kA} and then superimpose the resulting quiver with the following quiver 
\be \label{E8SOSp}
\rnode{}{2} - \bnode{}{2}-\rnode{}{4} - \bnode{}{4} - \rnode{}{6} - \bnode{}{6} - \rnode{\bver{}{2}}{8} - \bnode{}{6}- \rnode{}{6} - \bnode{}{4}-\rnode{}{4}  - \bnode{}{2}- \rnode{}{2}
\ee
where the leftmost red node in \eref{E8SOSp} is aligned with the leftmost node over the brace in \eref{mirrUSp2km8w2kA}.  Note that as a result of the superposition, each $SO(\text{odd})$ group over the brace in \eref{mirrUSp2km8w2kA} becomes an $SO(\text{even})$ group, with the rank increased according to the red nodes in \eref{E8SOSp}.   For the symplectic groups, one simply adds up their rank in a straightforward manner.  We thus arrive at \eref{SOSpmirr}, as required.  

Let us now consider quiver \eref{E8SOSp} in more detail.  This star-shaped quiver is, in fact, \eref{SOSpmirr} with $k=4$.  It is formed by gluing $T_{[1^8]}(SO(8))$, $T_{[1^8]}(SO(8))$ and $T_{[5,3]}(SO(8))$ together and gauging the common flavour symmetry $SO(8)$.  Hence it is a mirror theory of the $S^1$ compactification of the class $\mathsf{S}$ theory associated with a sphere of type $SO(8)$ with punctures $[1^8]$, $[1^8]$, $[5,3]$.  This theory of class $\mathsf{S}$ is indeed identified with the rank-1 $E_8$ SCFT (see \cite[p. 26]{Chacaltana:2011ze}), confirming our expectations that the transverse slice for any $k>3$ is the closure of the minimal nilpotent orbit of $E_8$.

\subsection{Relations at infinite coupling}
The ring of BPS operators on the Higgs branch at finite coupling has an $SO(4k)$ adjoint valued generator $M^{ij}$, with $i,j=1,\ldots, 4k$, as in \eref{finiterel}, and at infinite coupling admits a new generator $S_\alpha$ which transforms in the spinor representation of $SO(4k)$, hence $\alpha=1\ldots2^{2k-1}$. We recall that the $SU(2)_R$ representations of $M$ and $S$ assign the weights 2 and $k-2$, respectively.
At finite coupling we have two relations at weight 4, taking from \eref{finiterel},
\be
M^{ij} M^{kl} \delta_{jk} = \frac{1}{4k}\delta^{il} \Tr(M^2)
\ee
This relation is not modified at infinite coupling.
Next the relation
\be
\Tr(M^2) = 0 
\ee
at finite coupling is no longer valid at infinite coupling, and gets corrected as is clear from the relations at weight $2k-4$.
At infinite coupling we get a new relation at weight $k$,
\be
M^{ij} \gamma^{j}_{\alpha \dot \alpha} S_\alpha = 0,
\ee
where the $\gamma$ matrices have spinor indices which are contracted with a $\delta$ symbol for $k$ even and with an $\epsilon$ symbol for $k$ odd.
At weight $2k-6$ the classical rank constraint
\be
M^{[i_1 i_2} \cdots M^{i_{2k-7} i_{2k-6}]} = 0
\ee
is no longer valid and gets replaced at infinite coupling by relations at order $2k-4$ which read
\be
\Tr(M^2)^{\frac{k-2j-2}{2}}  
M^{[i_1 i_2} \cdots M^{i_{4j-1} i_{4j}]} = S_\alpha \gamma^{i_1\cdots i_{4j}}_{\alpha \beta}S_{\beta}~ \quad \text{for $k$ even}
\ee
with $j=0,1,\ldots, (k-2)/2$, and
\be
\Tr(M^2)^{\frac{k-2j-3}{2}}  
M^{[i_1 i_2} \cdots M^{i_{4j+1} i_{4j+2}]} = S_\alpha \gamma^{i_1\cdots i_{4j+2}}_{\alpha \beta}S_{\beta}~ \quad \text{for $k$ odd}
\ee
with $j=0,1,\ldots, (k-3)/2$.  Note that since a higher symmetric power of the adjoint representation of $SO(4k)$ contains non-trivial multiplicities, precise expressions for the relations can be complicated and difficult to write down.  Here we only report the simplest expressions of the possible relations which are general for any $k$. We leave a more complicated and careful analysis of the relations for future work. 
\\~\\
For $k=4$ these relations take the form
\bea
\Tr(M^2) &= S_\alpha \delta_{\alpha \beta} S_\beta \\
M^{[i_1 i_2}M^{i_3 i_4]} &= S_\alpha \gamma^{i_1i_2i_3i_4}_{\alpha \beta} S_\beta
\eea
For $k=5$ these relations take the form
 \bea
\Tr(M^2) M^{i_1 i_2}&= S_\alpha \gamma^{i_1 i_2}_{\alpha \beta} S_\beta \\
M^{[i_1 i_2}M^{i_3 i_4}M^{i_5 i_6]} &= S_\alpha \gamma^{i_1i_2i_3i_4i_5i_6}_{\alpha \beta} S_\beta
\eea
For $k=6$ these relations take the form
\bea
\Tr(M^2)^2 &= S_\alpha \delta_{\alpha \beta} S_\beta \\
\Tr(M^2) M^{[i_1 i_2}M^{i_3 i_4]} &= S_\alpha \gamma^{i_1i_2i_3i_4}_{\alpha \beta} S_\beta \\
M^{[i_1 i_2}\cdots M^{i_7 i_8]} &= S_\alpha \gamma^{i_1\cdots i_8}_{\alpha \beta} S_\beta
\eea
For $k=7$ these relations take the form
 \bea
\Tr(M^2)^2 M^{i_1 i_2}&= S_\alpha \gamma^{i_1 i_2}_{\alpha \beta} S_\beta \\
\Tr(M^2) M^{[i_1 i_2}M^{i_3 i_4}M^{i_5 i_6]} &= S_\alpha \gamma^{i_1i_2i_3i_4i_5i_6}_{\alpha \beta} S_\beta \\
M^{[i_1 i_2}\cdots M^{i_9 i_{10}]} &= S_\alpha \gamma^{i_1\cdots i_{10}}_{\alpha \beta} S_\beta
\eea
\section{Multiple M5-branes on $\BC^2/D_k$}  \label{sec:M5Dk}
Let us now consider the worldvolume of $N$ M5-branes on $\BC^2/D_k$.  This is a 6d $\CN=(1,0)$ with $2N-1$ tensor multiplets with gauge groups and hypermultiplets denoted in the following quivers  \cite{Intriligator:1997kq,  Brunner:1997gk,Blum:1997mm, Intriligator:1997dh, Brunner:1997gf, Hanany:1997gh, Ferrara:1998vf, DelZotto:2014hpa}:
\be \label{multM5}
\sqrnode{}{2k} - \underbrace{\bnode{}{2k-8} - \rnode{}{2k} - \bnode{}{2k-8}  - \cdots -  \bnode{}{2k-8}}_{\substack{N\,\,\text{blue nodes} \\ N-1\,\,\text{red nodes}}}- \sqrnode{}{2k}
\ee
The Higgs branch dimension at infinite coupling is given by \cite[(4.29)]{Mekareeya:2017sqh}:
\be \label{infdimHNM5}
\begin{split}
	\dim_\BH \,\, \CH_\infty [\eref{multM5}]  &= 29N_{T \rightarrow H} +n_H-n_V \\
	&= 29N + k(2k-8)(2N) \\
	& \qquad -\frac{1}{2}(2k-8)(2k-7)N - \frac{1}{2}(2k)(2k-1)(N-1) \\
	& = N+ \frac{1}{2} (2k)(2k-1) = N + \dim\,(SO(2k))~,
\end{split}
\ee
where it was pointed out in \cite[(5.1)]{Mekareeya:2017sqh} that only $N_{T \rightarrow H} = N$, out of $2N-1$ tensor multiplets, turn into hypermultiplets at the infinite coupling point, as is of course obvious from the brane picture of \cite{Hanany:1997gh}, since the number of physical NS5 branes is $N$.

As shown in \cite{Ohmori:2014kda}, the anomaly polynomial for the theory \eref{multM5}, with the centre of mass mode subtracted, takes the same form as \eref{anompoly} but with
\bea \label{anompolyDk}
\alpha &= \frac{1}{24} |\Gamma_{D_k}|^2 N^3 - \frac{1}{12} N \left[ |\Gamma_{D_k}| (k+1) -1 \right] + \frac{1}{24} [\dim(SO(2k)) -1]~,\\
\beta &= \frac{N}{48} \left[2-  |\Gamma_{D_k}|(k+1) \right] +  \frac{1}{48} [\dim(SO(2k)) -1]~, \\
\gamma &= \frac{1}{5760} \left[ 30(N-1) +7 \left\{ \dim(SO(2k)) +1 \right \} \right] \\
\delta &= -\frac{1}{1440} \left[ 30(N-1) +  \dim(SO(2k)) +1 \right]~,
\eea
while the other coefficients are the same as in \eref{anompoly}.

\subsection{$T^2$ compactification of the $6d$ theory}

The $T^2$ compactification of the SCFT at infinite coupling of $\eref{multM5}$, assuming $N>k$, gives the following system in four dimensions (see \cite{Ohmori:2015pia} and \cite[Eq.~(3.3.58)]{Ohmorithesis2015}):
\be \label{confmattDkn}
\frac{\mathsf{S} \langle T^2 \rangle_{SU(2N)}\{ \underline{TM},\underline{TM}, \underline{TM}, {\blue \underline{O}_k} \} \times \mathsf{S} \langle S^2 \rangle_{SO(2k)}\{ {\purple [1^{2k}]}, [1^{2k}], [1^{2k}] ] \}}{\diag({\blue SO(2k)} \times {\purple SO(2k)} )}
\ee
where 
\bi
\item $\mathsf{S} \langle T^2 \rangle_{SU(2N)}\{ \underline{TM},\underline{TM}, \underline{TM}, { \underline{O}_k} \}$ denotes a 4d class $\mathsf{S}$ theory of $SU(2N)$ type, whose Gaiotto curve is a torus with three minimal twisted punctures $\underline{TM}$ and a twisted puncture $\underline{O}_k$ with a symmetry $SO(2k)$.  

Using the notation of \cite{Chacaltana:2012ch} for the twisted $SU(2N)=A_{2N-1}$ class $\mathsf{S}$ theory,  the twisted puncture is labelled by a B-partition of $2N+1$, and the untwisted puncture is labelled by an ordinary partition of $2N$.  
\bi
\item The minimal twisted puncture $\underline{TM}$, whose flavour symmetry is trivial, is labelled by $[2N+1]$.  
\item The twisted puncture of $\underline{O}_k$, whose flavour symmetry is $SO(2k)$, is labelled by $[2(N-k)+1,1^{2k}]$.  
\ei
\item $\mathsf{S} \langle S^2 \rangle_{SO(2k)}\{ { [1^{2k}]}, [1^{2k}], [1^{2k}] ] \}$ denotes a 4d class $\mathsf{S}$ theory of $SO(2N)$ type, whose Gaiotto curve is a sphere with three maximal punctures $[1^{2k}]$, each with a symmetry $SO(2k)$.
\item The factor $\diag(SO(2k) \times SO(2k))$ in the denominator denotes the gauging of the diagonal subgroup of $SO(2k) \times SO(2k)$ coming from $\underline{O}_k$ and one of the $[1^{2k}]$ punctures.
\ei
The symmetry that is manifest in \eref{confmattDkn} is therefore $SO(2k) \times SO(2k)$.  As pointed out in \cite{Ohmori:2015pia, Ohmorithesis2015}, the gauge group $\diag(SO(2k) \times SO(2k))$ in \eref{confmattDkn} is infrared free.

\subsection{$T^3$ compactification of the $6d$ theory}
Let us now discuss the mirror theories of the $S^1$ compactification of \eref{confmattDkn}.  We first examine the mirror theories of the $S^1$ compactification of the theories of class $\mathsf{S}$ that appear in \eref{confmattDkn}.  The relevant 3d theories associated with the punctures $\underline{TM} =[2N+1]$, $\underline{O}_k= [2(N-k)+1,1^{2k}]$ and $[1^{2k}]$ are
\bea
T_{[2N+1]}(USp(2N)): &\quad \sqbnode{}{2N} \\
T_{[2(N-k)+1,1^{2k}]}(USp(2N)): &\quad \rnode{}{2} - \bnode{}{2}-\rnode{}{4} - \bnode{}{4} - \cdots -\rnode{}{2k-2}-\bnode{}{2k-2}-\rnode{}{2k}-\sqbnode{}{2N} \\
T_{[1^{2k}]} (SO(2k)): &\quad \rnode{}{2} - \bnode{}{2}-\rnode{}{4} - \bnode{}{4} - \cdots -\rnode{}{2k-2}-\bnode{}{2k-2}- \sqrnode{}{2k}
\eea

We conjecture that the mirror theory of the $S^1$ compactification of \eref{confmattDkn} is
\be \label{combinedquiv}
\begin{aligned}
	\rnode{}{2} - \bnode{}{2}-\rnode{}{4} - \bnode{}{4} - \cdots -\rnode{}{2k-2}-\bnode{}{2k-2}- \rnode{\overset{\overset{A'}{\cap}}{\bver{}{2N}}}{2k}- \bnode{}{2k-2}- \rnode{}{2k-2} - \cdots - \bnode{}{4}-\rnode{}{4}  - \bnode{}{2}- \rnode{}{2}
\end{aligned}
\ee
where $A'$ denotes the hypermultiplet in the traceless rank-2 antisymmetric representation $[0,1,0, \ldots,0]$ of $USp(2N)$.  Let us compute the Higgs and Coulomb branch dimensions of \eref{combinedquiv}.
\bea
\dim_\BH \,\CC[\eref{combinedquiv}] &= 2 \times 2 \left( \sum_{j=1}^{k-1} j\right)+k  + N  = N+\frac{1}{2}(2k)(2k+1) = \eref{infdimHNM5} ~, \\
\dim_\BH \,\CH[\eref{combinedquiv}] &= 2 \left[ \sum_{j=1}^{k-1} \frac{1}{2}(2j)(2j)  +\sum_{j=1}^{k-1} \frac{1}{2}(2j)(2j+2)   \right] + \frac{1}{2}(2N)(2k)  \nn\\
& + \left[ \frac{1}{2}(2N)(2N-1) -1\right] -2 \left[ \sum_{j=1}^{k-1} \frac{1}{2}(2j)(2j-1)+\sum_{j=1}^{k-1} \frac{1}{2}(2j)(2j+1)  \right] \nn \\ 
& - \frac{1}{2}(2k)(2k-1) -\frac{1}{2}(2N)(2N+1) \nn \\
&=  2 kN - 2N -k-1~.
\eea
It should be noted that 
\be
\begin{split}
	& n_T [\eref{multM5}] + (\text{the total rank of the gauge groups in \eref{multM5}}) \\
	&= (2N-1)+N(k-4)+(N-1)k \\
	&= 2kN -2N-k -1\\
	&= \dim_\BH \,\CH[\eref{combinedquiv}] ~.
\end{split}
\ee
This is to be expected as a mirror theory of the $T^3$ theory of the 6d theory \eref{multM5}.
Moreover, for $N=1$, the field $A'$ disappears and we recover \eref{SOSpmirr}.

For $k=3$, the $\BC^2/D_3$ singularity is in fact $\BC^2/\BZ_4$ singularity.  The  The HWG for the Coulomb branch of \eref{combinedquiv} for $N=2$ can be obtained from \cite[(2.29)]{Hanany:2018vph}. For reference, we report the result here:
\be
\begin{split}
&\text{HWG of $\CC[\eref{combinedquiv}]_{N=2,k=3}$}  \\
&= \PE \left[ \mu _1 \mu _7 t^2 +\left(\mu _4+\mu _2 \mu _6+1\right) t^4 +\left(\mu _4+\mu _3 \mu _5\right) t^6 +\mu _4^2 t^8 -\mu _4^2 t^{12} \right]
\end{split}
\ee
where $\mu_i$ are highest weight fugacities for $SU(8)$.  Since the computation of the HWG for a general $k$ and $N$ is technically challenging due to the Hilbert series of the $T_{[1^{2k}]}(SO(2k))$ tails, we leave it for future work.

\section{Conclusion and further developments}
In this paper, we study the Higgs branches of theories on M5-branes on $\BC^2/D_k$ at finite and at infinite coupling. The extra massless states that arise at the infinite coupling point are analysed in detail for the case of a single M5 brane using the Coulomb branch of the corresponding 3d gauge theories.  The small instanton transition that gives rise to such massless states can be realised in an elegant way using a superposition of two 3d quivers, shown in \eref{superposition}.  This process is referred to as the generalised Kraft--Procesi transition. Finally, we also propose a 3d quiver theory \eref{combinedquiv} whose Coulomb branch describes the Higgs branch of the theory on $N$ M5-branes on $\BC^2/D_k$ at infinite coupling.

After the appearance of the first version of this paper on {\tt arXiv}, there have been further developments along the direction of this paper.  For example, a 3d quiver theory \eref{combinedquiv} whose Coulomb branch describes the Higgs branch of the theory on $N$ M5-branes on $\BC^2/\BZ_k$ at infinite coupling was proposed in \cite[(2.7)]{Hanany:2018vph} via gauging of the discrete symmetry:
\be \label{infiniteZk}
\node{}{1}- \node{}{2}-\cdots -\node{}{k-1}-\node{\overset{\cap}{\ver{}{\,\,N}}}{k}-\node{}{k-1}-\cdots - \node{}{2}-\node{}{1}
\ee
where the node with a label $m$ denotes a $U(m)$ gauge group, the notation $\cap$ denotes an adjoint hypermultiplet under the $U(N)$ gauge group, and an overall $U(1)$ is modded out from this quiver.  This is indeed a generalisation of \eref{combinedquiv} in the following sense. Both \eref{combinedquiv} and \eref{infiniteZk} are star-shaped quivers consisting of three legs, where the two long legs come from the $T(G)$ theory \cite{Gaiotto:2008ak}, with $G$ being $U(k)$ or $O(2k)$ for the cases of $\BZ_k$ and $D_k$ orbifolds respectively, and the other leg comes from attaching a gauge group of rank $N$ to the central node $G$ with an extra hypermultiplet in an appropriate representation (adjoint and rank-two-antisymmetric-traceless representations for $\BZ_k$ and $D_k$ respectively).  

An immediate generalisation of \eref{combinedquiv} and \eref{infiniteZk} is to obtain 3d quiver whose Coulomb branch describes the infinite coupling Higgs branch of the T-brane theories $\CT_G{(\{Y_L, Y_R\},N-1)}$\footnote{Here we use the same notation as in \cite{Mekareeya:2017sqh}.}, which constitute another class of models by turning on the nilpotent orbits $Y_L$ and $Y_R$ of $G$ \cite{Heckman:2016ssk}.  We conjecture that this can be achieved by simply replacing the two long legs corresponding to $T(G)$ by $T_{Y_L}(G)$ and  $T_{Y_R}(G)$.  In particular, for $G=SU(k)$, the corresponding 3d quiver is
\be
\frac{T_{Y_L}(SU(k))  \times \sqnode{\overset{\cap}{\ver{}{\,\,N}}}{k} \times T_{Y_R}(SU(k)) }{U(k)/U(1)}~,
\ee
and for $G=SO(2k)$, the corresponding 3d quiver is
\be
\frac{T_{Y_L}(SO(k))  \times  \sqrnode{\overset{\overset{A'}{\cap}}{\bver{}{2N}}}{2k} \times T_{Y_R}(SO(k)) }{SO(k)/\BZ_2}~.
\ee
One can check that the Coulomb branch dimension of these 3d theories is
\be
\begin{split}
&\left[ \frac{1}{2}\left(\dim(G)-\mathrm{rank}(G) \right) -\dim_{\BH} Y_L\right] +N \\
&+ \left[ \frac{1}{2}\left(\dim(G)-\mathrm{rank}(G) \right) -\dim_{\BH} Y_R \right] + \mathrm{rank}(G)  \\
&= N+ \dim(G) - \dim_{\BH} Y_L -\dim_{\BH} Y_R
\end{split}
\ee
in agreement with the infinite coupling Higgs branch dimension of the corresponding 6d T-brane theory \cite[(1.2)]{Mekareeya:2017sqh}. 

It would be interesting to generalise such description to the $E$-type singularities.  To the best of our knowledge, the $T(E_{6,7,8})$ theories do not admit a quiver description.  Furthermore, it remains to identify an appropriate node of rank $N$, as well as the ``hypermultiplet'' that transform under such a node.  We leave this for future work.

\acknowledgments
A.~H.~ and N.~M.~ gratefully acknowledge the Pollica Summer Workshop 2017 (partly supported by the ERC STG grant 306260) for their kind hospitality, where this project was initiated, and the Simons Summer Workshop 2017, where progress on this project has been made.
A.~H.~ would like to thank Vishnu Jejjala and the National Institute of Theoretical Physics at the University of the Witwatersrand (WITS) for their kind hospitality during the last stages of this project.
The research of N.~M.~ is supported by the INFN.
A.~H.~ is supported in part by an STFC Consolidated Grant ST/J0003533/1, and an EPSRC Programme Grant EP/K034456/1.

\bibliographystyle{ytphys}
\bibliography{ref,refA}

\end{document}